%% file: CSCW17-WorkshopPaper.tex
\documentclass{sigchi-ext}
\usepackage[T1]{fontenc}
\usepackage{textcomp}
\usepackage[scaled=.92]{helvet} % for proper fonts
\usepackage{graphicx} % for EPS use the graphics package instead
\usepackage{balance}  % for useful for balancing the last columns
\usepackage{booktabs} % for pretty table rules
\usepackage{ccicons}  % for Creative Commons citation icons
\usepackage{ragged2e} % for tighter hyphenation
\usepackage{stmaryrd} % for extra latex symbols (lighting)
\usepackage[usenames,dvipsnames]{xcolor}
\usepackage[normalem]{ulem}
\usepackage{seqsplit}

\title{How Software Developers Mitigate Collaboration Friction with Chatbots}
\numberofauthors{3}
\author{%
  \alignauthor{%
    \textbf{Carlene Lebeuf}\\
    \affaddr{University of Victoria} \\
    \affaddr{Victoria, BC, Canada} \\
    \affaddr{clebeuf@uvic.ca} }\alignauthor{%
    \textbf{Alexey Zagalsky}\\
    \affaddr{University of Victoria}\\
    \affaddr{Victoria, BC, Canada}\\
    \email{alexeyza@uvic.ca} } \vfil \alignauthor{%
    \textbf{Margaret-Anne Storey}\\
    \affaddr{University of Victoria} \\
    \affaddr{Victoria, BC, Canada} \\
    \affaddr{mstorey@uvic.ca} }}

\def\plaintitle{How Software Developers Mitigate Collaborative Friction Points with Chatbots} \def\plainauthor{Carlene Lebeuf, Margaret-Anne Storey, Alexey Zagalsky}
\def\plainkeywords{Software Development, Collaboration, Chatbots, Bots, Conversational Interfaces, HCI, Socio-Technical Systems}

\hypersetup{%
  pdftitle={\plaintitle}, pdfauthor={\plainauthor},
  pdfkeywords={\plainkeywords}, }

\begin{document}

\maketitle

% Uncomment to disable hyphenation (not recommended)
% https://twitter.com/anjirokhan/status/546046683331973120
\RaggedRight{} 

% Do not change the page size or page settings.
\begin{abstract}
Modern software developers rely on an extensive set of social media tools and communication channels. The adoption of team communication platforms has led to the emergence of conversation-based tools and integrations, many of which are chatbots. Understanding how software developers manage their complex constellation of collaborators in conjunction with the practices and tools they use can bring valuable insights into socio-technical collaborative work in software development and other knowledge work domains.

In this paper, we explore how chatbots can help reduce the friction points software developers face when working collaboratively. Using a socio-technical model for collaborative work, we identify three main areas for conflict: friction stemming from team interactions with each other, an individual's interactions with technology, and team interactions with technology. Finally, we provide a set of open questions for discussion within the research community.

\end{abstract}

\keywords{\plainkeywords}

\category{H.5.2}{User Interfaces}{Natural Language}

\input{sections/introduction}

\input{sections/friction_points_summary}
\input{sections/background}
\input{sections/friction_points}

\input{sections/discussion-points}

\balance{} 

\bibliographystyle{SIGCHI-Reference-Format}
\bibliography{references}

\end{document}

%% file: sections/introduction.tex
%!TEX root = ../CSCW17-WorkshopPaper.tex

\section{Introduction}

\begin{marginfigure}[0pt]
  \begin{minipage}{\marginparwidth}
    \centering
    \includegraphics[width=0.9\marginparwidth]{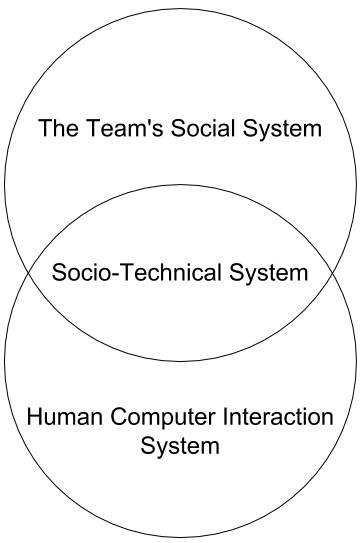}
    \caption{A simplified view of the intersection between domain models: the team's social system consists of collaborative work between people; the human computer interaction system focuses on individuals' interactions; and the socio-technical systems lies in the intersection of both.}~\label{fig:systems}
  \end{minipage}
\end{marginfigure}

Many software projects have adopted team communication and collaboration tools such as Slack, Hipchat, Microsoft Teams, and Flowdock \cite{Storey7498605}. These IRC-like communication tools serve as a \textit{central hub} for the organization, providing a fully searchable interface with persistent chat rooms, private group messages, and direct messages. Not surprisingly, distributed software teams rely on these team messaging platforms for the majority of their team communication, but collocated teams also use such systems to facilitate collaboration and promote a participatory culture. While software developers are at the forefront of the design and exploration of new tools---they can manipulate tools to fit their needs---other knowledge worker domains (e.g., journalism) are also adopting popular collaborative software development tools. Studying how software developers manage the complex constellation of collaborators in conjunction with their tools and practices provides valuable insights into understanding socio-technical collaborative work.

This widespread adoption of team-based communication platforms provides a breeding ground for new conversation-based tools and integrations. Many of these integrations are in the form of conversational bots (also referred to as chatbots). Although bots have been around for many years, the ability to easily integrate them with modern communication tools and access various APIs and data sources has created a recent explosion of new bots. These chatbots can be powered by rules, machine learning, or artificial intelligence, allowing users to interact via text or even spoken words (e.g., \texttt{MyVoiceBot} connects Amazon's Alexa to other chatbots inside Hipchat, giving the user voice control over typical text-based interactions). 

Developers have been adopting these chatbots at a dizzying pace to support and fill many of the roles that software developers have traditionally filled: maintaining code quality, preforming testing, conducting development operations, supporting customers, and creating documentation \cite{Storey:2016:DDP:2950290.2983989}. Yet the research community struggles to understand how chatbots are supporting (or hindering) collaborative work, why some bots are more useful than others, and what risks they introduce.

% In our previous studies, we came to realize that developers face many friction points not only in using tools, but also in working with eachother. We can use these friction points, what we have observed is that we can use these chatbots, or developers are using these chatbots to address these friction points. So in this paper we talk about the friction points but also the chatbots that are being used to address them.

Based on our past research, we realize that developers face friction points not only when using tools, but also when working with each other \cite{Storey:2012:ESP:2664446.2664469,Storey7498605,Arciniegas-Mendez:2017}. Recently, developers have been creating their own breed of chatbots to address the collaboration needs these friction points expose. 

In this paper, we explore how chatbots help reduce collaboration friction points in software development. Specifically, we aim to answer the following research question: 

\textbf{RQ: What collaboration friction points can chatbots help mitigate in software development, and how?}

Through a study of existing literature, an analysis of popular chatbots, our past studies~\cite{Storey7498605,Storey:2012:ESP:2664446.2664469,Storey:2016:DDP:2950290.2983989}, and our personal software development experiences, we synthesized a set of friction points software developers face when working collaboratively. This is not a comprehensive list of all possible areas of conflict, but rather a curated list focusing on common collaboration friction points. 

We classify and present three categories of collaborative friction points inspired by the model of socio-technical systems~\cite{whitworth2009social}. Figure~\ref{fig:systems} shows a simplified view of the relationships between the systems. For each category, we found various chatbots being used by software teams, which we believe highlights the growing challenges in modern collaborative work environments.

We conclude the paper with a set of research questions to spark discussion on the future use of chatbots in software development.

%% file: sections/friction_points_summary.tex
%!TEX root = ../CSCW17-WorkshopPaper.tex

\marginpar{%
  \fbox{%
    \begin{minipage}{0.925\marginparwidth}
    	{\centering
		  \textbf{Collaboration Friction Points in Software Development}\par
		}

		\textit{ \\Team Interactions}
		\begin{itemize}[topsep=0pt,itemsep=0pt,parsep=0pt,partopsep=20pt,leftmargin=10pt]
			\item[$\lightning$] Understanding team members' roles and expertise
			\item[$\lightning$] Adhering to team procedures and agreements
			\item[$\lightning$] Understanding and working towards team goals
			\item[$\lightning$] Coordinating team activities
			\item[$\lightning$] Managing trust and team cooperation
		\end{itemize}

		\textit{ \\Individuals $\Leftrightarrow$ Technology}
		\begin{itemize}[topsep=0pt,itemsep=0pt,parsep=0pt,partopsep=20pt,leftmargin=10pt]
			\item[$\lightning$] Distracting and interruptive technologies
			\item[$\lightning$] Maintaining awareness of new technologies
			\item[$\lightning$] Understanding channel affordances
		\end{itemize}

		\textit{ \\Teams $\Leftrightarrow$ Technology}
		\begin{itemize}[topsep=0pt,itemsep=0pt,parsep=0pt,partopsep=20pt,leftmargin=10pt]
			\item[$\lightning$] Information fragmentation and overload
			\item[$\lightning$] Adopting and understanding tool usage in the team's context
			\item[$\lightning$] Maintaining awareness of project activities
			\item[$\lightning$] Inadequate collaboration tooling
			\item[$\lightning$] Miscommunication on text-based channels
		\end{itemize}
    \end{minipage}}\label{sec:sidebar} }

%% file: sections/background.tex
%!TEX root = ../CSCW17-WorkshopPaper.tex

\section{Background}

% The background section of the paper. Talk about difference CSCW theories such as regulation, awareness, alignment, transparency, etc. 

We define a collaboration friction point ($\lightning$) as any resistance or conflict arising from a team's joint processes. These friction points may stem from the developers' mandatory collaborative needs not being fully satisfied.

Since friction can occur at various levels in a team's collaborative activities, we explored Whitworth's Model of Socio-Technical Systems \cite{whitworth2009social} as a way to classify the types of friction developers experience.
A socio-technical system is a complex social system evolving around a technical base, which includes the interplay of human, social, environmental, and technological factors.
The lowest level contains the \textbf{hardware system}, or the collection of \textit{physical} parts.   
The \textbf{software system} emerges from the hardware system and is based on \textit{information}, the exchange of data, and code. 
With the addition of \textit{personal} exchanges between the software and a human, the \textbf{human-computer interaction (HCI) system} is activated. 
The \textbf{socio-technical (ST) system}, the highest level, emerges from the HCI system with the addition of \textit{community} exchanges.

We expanded the Model of Socio-Technical Systems (Figure \ref{fig:STS}) to encapsulate the societal and team social systems. We identified that collaboration breakdowns can occur at three levels: \textbf{$\lightning$ team interactions} (team's social system), \textbf{$\lightning$ individuals' interactions with technology} (HCI system), and \textbf{$\lightning$ the team's interactions with technology} (ST system).  
The sidebar presents a high-level summary of the friction points, classified by their interaction level.

In the following sections, we detail the levels of collaborative friction experienced by developers, summarize select collaboration friction points, and then suggest popular \texttt{chatbots} that we believe can help mitigate the detrimental effects of this friction. 

%% file: sections/friction_points.tex
%!TEX root = ../CSCW17-WorkshopPaper.tex

\begin{figure}[t]
  \includegraphics[width=\columnwidth]{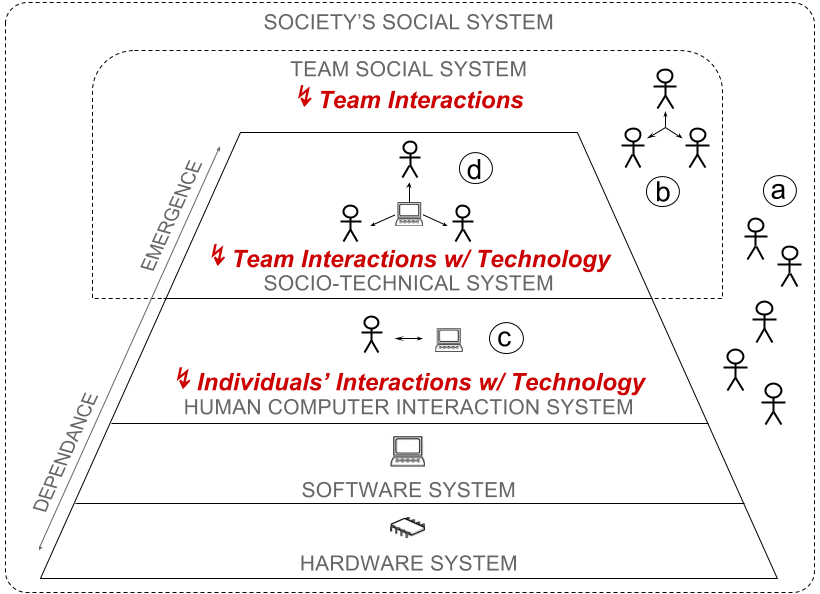}
  \caption{Socio-Technical Model for Collaborative Software Development including the (a) society's social system, (b) team's social system, and 3 categories of friction points: (b) $\lightning$ team's interactions, (c) $\lightning$ individual's interactions with technology, and (d) $\lightning$ team's interactions with technology.}~\label{fig:STS}
  \vspace{-12pt}
\end{figure}

%%%%%%%%%%%%%%%%%%%%%%%%%%%%%%%%%%%%%%%%%%%%%%%%%%%%%%%%%%%%%%%%%%%%%%%%%%%%%%%%%%%%%%%%%%%%%%%%%%%%%%%%%%%%%%
\section{Friction in Team Interactions}

This type of collaborative friction occurs as a result of interactions within the team's social system as they work together to realize and achieve shared goals.
Although technology may facilitate these interactions, the friction stems from collaborative processes and would occur regardless of the use of technology.

\subsection{\texorpdfstring{$\lightning$}{!}\textbf{Understanding team members' roles and expertise.}}
Understanding team members' roles and responsibilities can be difficult in large or distributed teams.
Developers often experience difficulty locating the best team member for mentorship \cite{Steinmacher:2012:RMS:2666719.2666734}, assistance on tasks requiring special expertise \cite{Moraes:2010:REU:1808920.1808929}, or help when experience with certain aspects of the code base is needed, e.g., bug fixes \cite{Anvik:2006:FTB:1134285.1134336}.

\begin{marginfigure}[-15pt]
  \begin{minipage}{\marginparwidth}
    \centering
    \includegraphics[width=1\marginparwidth]{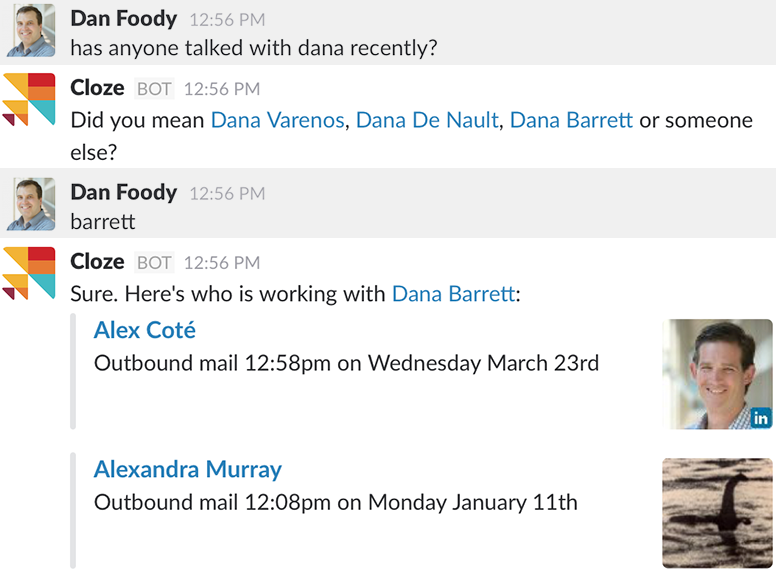}
    \caption{\texttt{Cloze} slackbot. Photo: \url{https://www.cloze.com/}
    }~\label{fig:cloze}
  \end{minipage}
\end{marginfigure}

To help maintain awareness of teammates' roles, developers use \texttt{WhoBot} and \texttt{Cloze}. Asking \texttt{WhoBot} "who knows about ..." returns a list of team members that frequently mentioned the topic, and \texttt{Cloze} consolidates everything you need to know about your contacts in one place: background information, a summary of interactions, and any follow-up items or notes.

\begin{marginfigure}[0pt]
  \begin{minipage}{\marginparwidth}
    \centering
    \includegraphics[width=1\marginparwidth]{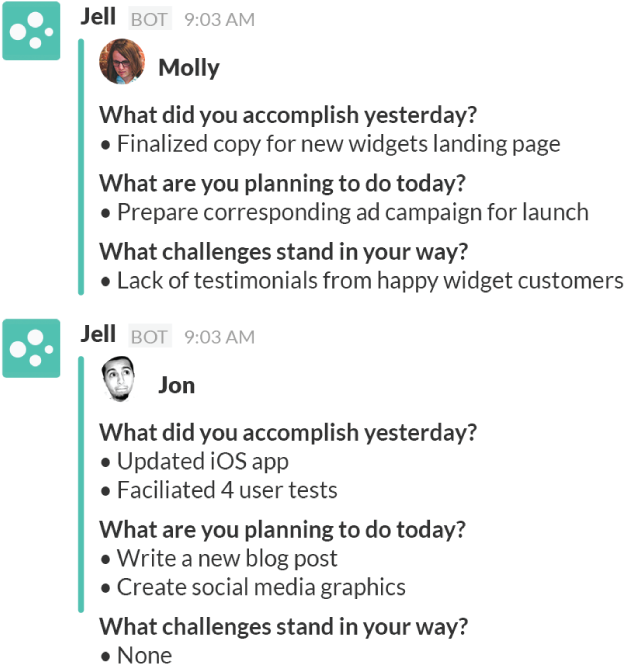}
    \caption{\texttt{Jell} slackbot. Photo: \url{https://jell.com/}
    }~\label{fig:jell}
  \end{minipage}
\end{marginfigure}

\begin{marginfigure}[0pt]
  \begin{minipage}{\marginparwidth}
    \centering
    \includegraphics[width=1\marginparwidth]{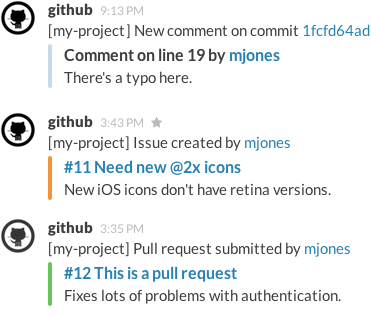}
    \caption{\texttt{Github} slackbot. Photo: \url{https://twitter.com/SlackHQ/status/448572066808090624}
    }~\label{fig:github}
  \end{minipage}
\end{marginfigure}

% MOVE TO DISCUSSION?
% Inspiration for chatbots to help connect team members can also come from other industries.  Although not currently targeted at software development, Slack's MentorBot allowing users ask questions to the chatbot and get advice from successful entrepreneurs in return.
% One could imagine a similar experience created to connect software developers to connect with experts both within, and outside their teams.  
% Following along this line, the slackbot Drift\footnote{https://www.drift.com/slack/}, helps direct customers to the correct team member to help them with their issue.  
% One could easily imagine a similar chatbot that directs developers to other available member of their projects, particularly sub-teams with which they may not have frequent contact such as testers or devOps. 
% These chatbots could be expanded to include context aware recommendation systems \cite{Moraes:2010:REU:1808920.1808929}, providing a powerful, targeted way to ask for help directly in the teams communication tool.

\subsection{\texorpdfstring{$\lightning$}{!}\textbf{Adhering to team's procedures and agreements.}}
Scripts, often referred to as working agreements, are a set of rules or guidelines that teams agree to follow to make themselves---and the team as a whole---more efficient \cite{Arciniegas-Mendez:2017}.  
Team members often forget or are unaware of the correct sequence of steps when completing tasks, particularly new members joining the team \cite{Arciniegas-Mendez:2017}.

A development team at a large software company uses \texttt{Slackbot} to answer users' queries regarding the correct procedures for development activities, such as merging feature branches into the main code base. 
They also use \texttt{Slackbot} to monitor system status (e.g., website outages), notify the team, and provide a set of next steps.

% MOVE TO DISCUSSION?
% Engineering Health Checker, a virtual 'wall of shame' of developers who recently introduced heath issues in the code could be adapted into a chatbot supporting shared- regulation

\subsection{\texorpdfstring{$\lightning$}{!}\textbf{Understanding and working towards team goals.}}
To collaborate successfully, all team members must understand the team's goals and the actions needed to achieve them \cite{Arciniegas-Mendez:2017}. They must also maintain awareness of the project's overall status by communicating what they are working on, their progress, and blockers they face \cite{Arciniegas-Mendez:2017}. 

To help the team maintain awareness of each other's activities, developers use chatbots like \texttt{WorkingOn} to broadcast real-time status updates to the group, and \texttt{Jell} to share their accomplishments, ongoing tasks, and blockers. 
% \texttt{Screenfully} provides similar functionality with the addition of visualizations of the team's performance.

\subsection{\texorpdfstring{$\lightning$}{!}\textbf{Coordinating team activities.}}
For large or distributed teams, coordinating multiple meetings, deadlines, and other activities can be a complex task. However, it can be even more difficult to ensure everyone is made aware of and remembers activities.  

To help find a meeting time that works for everyone, developers turn to \texttt{Meekan}, a chatbot that searches through everyone's calendars, returns the times with the least conflicts, and allows team members to vote on their preferred options. 
Before a meeting, the \texttt{Solid} chatbot sends detailed reminders to team members. 
During the meeting, \texttt{Solid} keeps track of the time remaining and any incomplete tasks. After the meeting, \texttt{Solid} sends attendees the meeting's outcomes and any tasks they were assigned.
% Team's already using \texttt{Google Calendar} can take advantage of their slackbot integration to add items to their calendars and receive reminders for upcoming meetings.

\subsection{\texorpdfstring{$\lightning$}{!}\textbf{Managing trust and team cooperation.}}
Collaborating with others is difficult; some team members may have poor attitudes or lack the willingness to participate in a collaborative manner \cite{Storey7498605}, while others may take a while to warm up to and begin trusting their teammates \cite{Al-Ani:2013:GDS:2441776.2441840}.

\texttt{Oskar} asks and shares how individuals are feeling to prevent isolation and allow teammates to offer support.
To monitor team morale, \texttt{Ava Bot} privately checks in with team members to see how things are going and raises issues to management when appropriate.

\section{Friction in Interactions with Technology}
Two categories of friction points stem directly from technologies not adequately supporting software developers' collaboration needs: friction resulting from \textbf{individual} and \textbf{team} interactions with technology.

Friction resulting from \textbf{individuals' interactions} with technology arises when an \textit{individual's needs} are not fulfilled by the technologies they are using, which impedes their ability to collaborate effectively.
Although an overview of these friction points is provided in the sidebar, for the sake of brevity, we focus on team interactions with technology. 

Friction resulting from \textbf{team interactions} with technology arises when the \textit{community's needs} are not fully satisfied by the technologies they are using.

\input{sections/chatbot_sidebar}
\subsection{\texorpdfstring{$\lightning$}{!}\textbf{Information overload.}}
Attempting to enable developers to work more productively, teams often adopt more tools in their workflow, resulting in information and channel overload \cite{Storey7498605}. As the team's knowledge gets spread over the growing number of channels, issues with information fragmentation and quality begin to emerge \cite{Storey7498605}.

To deal with this ever growing flow of knowledge artifacts, developers use chatbot integrations for many of their everyday tools \cite{Storey:2012:ESP:2664446.2664469,Calefato2016AHM,Storey:2016:DDP:2950290.2983989}.
Chatbots also provide them curated overviews of channels they may not be actively following: \texttt{Digest.ai} creates daily recaps of team discussions, and \texttt{TLDR} generates summaries for long messages.

\subsection{\texorpdfstring{$\lightning$}{!}\textbf{Adopting \& understanding tool usage in the team's context.}}
Friction occurs when team members refuse to adopt the tools required for their job, which is often due to a lack of technical knowledge\cite{Kalliamvakou:2015:OSC:2818754.2818825}.
Even with successful adoption, team members still need to understand how to use these tools within their team's social context.

To help teams learn to use communication tools, \texttt{Slackbot} and \texttt{T-Bot} teach teammates how to preform common actions, such as creating a new channel. 
Developers also use chatbots to "bring technology into the conversation" and make complex tools or tasks (e.g., development operations) accessible to the entire team though text-based commands in their communication platforms\footnote{https://youtu.be/IhzxnY7FIvg}. 
% This allows team members without advanced expertise to perform simple automation, makes triggered actions visible to the entire team, and frees up domain experts' time.

% \subsection{\texorpdfstring{$\lightning$}{!}\textbf{Inadequate Collaboration Tooling.}}
% Developers have identified a lack of adequate tool support for sharing, explaining, collaborating, and receiving feedback on code \cite{Storey7498605}. 
% Many of the communication channels they are required to use for these tasks are actually adding friction, obstructing participatory development activities \cite{Storey7498605}.

% Chatbots needed...

\subsection{\texorpdfstring{$\lightning$}{!}\textbf{Maintaining awareness of the team's technology-dependent activities.}}
With the surge of social development tools, developers must maintain awareness and coordinate their development activities with those of their teammates \cite{Storey7498605}. 
For successful collaboration to occur, they need to understand how to use the tools within the context of others using them as well.

Chatbot-style integrations for existing tools like \texttt{GitHub}, \texttt{BitBucket}, and \texttt{Subversion} notify the team when changes are made to the codebase, helping developers maintain awareness of the team's collaborative activities.  

% \subsection{\texorpdfstring{$\lightning$}{!}\textbf{Miscommunication on text-based channels}}

% Chatbots needed...
% \carly{I wonder if this would be better to be one that we don't talk about} \cite{Storey7498605}

%% file: sections/chatbot_sidebar.tex
%!TEX root = ../CSCW17-WorkshopPaper.tex

\marginpar{%
	\vspace{-200pt}
  	\fbox{%

    \begin{minipage}{0.925\marginparwidth}

    	{\centering
		  \textbf{Glossary of Chatbots}\par
		}
		\texttt{ \\Ava Bot}\\ \url{http://zeal.technology}	

		\texttt{BitBucket}\\ \url{https://slack.com/apps/A0F7VRDPE-bitbucket}

		\texttt{Convergely}\\ \url{https://www.convergely.com}

		\texttt{Cloze}\\ \url{https://www.cloze.com/app/connect/slack}	

		\texttt{Digest.ai}\\ \url{https://digest.ai/} 	

		\texttt{Github}\\ \url{https://github.com/integrations/slack}

		\texttt{WhoBot}, \texttt{T-Bot}\\ \url{https://www.onmsft.com/news/microsoft-teams-introduces-t-bot}\\\url{-and-who-bot}		

		\texttt{Jell}\\ \url{https://jell.com/slack}

		\texttt{Knelfbot}\\ \url{http://www.knelf.com/slack.html}		

		\texttt{Meekan}\\ \url{https://meekan.com}	

		\texttt{MyVoiceBot}\\ \url{http://demo.softserveinc.com/}

		\texttt{Oskar}\\ \url{http://oskar.hanno.co}

		\texttt{Screenfully}\\ \url{http://screenful.com/guide/slack-integration}	

		\texttt{SlackBot}\\ \url{https://slack.com/apps/A0F81R8ET-slackbot}

		\texttt{Solid}\\ \url{https://getsolid.io/slackbot}

		\texttt{Subversion}\\ \url{https://slack.com/apps/A0F827LTA-subversion}

		\texttt{WorkingOn}\\ \url{https://www.workingon.co/integrations/slack}	

    \end{minipage}}\label{sec:sidebarChatbots} }

%% file: sections/discussion-points.tex
%!TEX root = ../CSCW17-WorkshopPaper.tex

\section{Discussion Points for the Workshop}

We believe that chatbots have immense potential for supporting developers' collaboration needs. 
However, we first need to understand the benefits and possible risks these new, virtual teammates are bringing to the software development teams that are so openly embracing them.

To conclude, we propose a set of research questions to spark discussion on the future of chatbots and their ability to facilitate collaboration in software development:

\begin{enumerate}[topsep=-5pt,parsep=5pt]
	\item How should we study chatbots? Can existing models and theories of collaboration help explain how chatbots are being used?
	\item What other collaborative friction points can be addressed with new or existing chatbots?
	\item With rapid progress being made in the fields of artificial intelligence, machine learning, and speech interfaces, how might this change the use of chatbots in the future?
	\item What risks are introduced by adopting chatbots in software development?
\end{enumerate}